\documentclass{revtex4}
\usepackage{amssymb}
\usepackage{graphicx}
\usepackage{lettrine} 
\usepackage{amsmath}

\begin{document}
\title{The effects of dissipation on topological mechanical systems}

\author{Ye Xiong$^{1,*,+}$, Tianxiang Wang$^1$, Peiqing Tong$^{1,2}$}
\affiliation{$^1$Department of Physics and Institute of Theoretical
  Physics
, Nanjing Normal University, Nanjing 210023,
P. R. China \\
$^2$Jiangsu Key Laboratory for Numerical Simulation of Large
Scale Complex Systems, Nanjing Normal University, Nanjing 210023,
P. R. China \\
$^*$xiongye@njnu.edu.cn \\
$^+$These authors contribute equally to this work}

\begin{abstract}

  We theoretically study the effects of isotropic dissipation in a
  topological mechanical system which is an analogue of Chern insulator
  in mechanical vibrational lattice. The global gauge invariance is
  still conserved in this system albeit it is destroyed by the
  dissipation in the quantum counterpart.  The chiral edge states in
  this system are therefore robust against strong dissipation. The
  dissipation also causes a dispersion of damping for the eigenstates.
  It will modify the equation of motion of a wave packet by an extra
  effective force. After taking into account the Berry curvature in the
  wave vector space, the trace of a free wave packet in the real space
  should be curved, feinting to break the Newton's first law.

\end{abstract}

\pacs{46.40.Cd, 63.22.-m, 03.65.Vf, 73.43.-f}


\maketitle

\section*{Introductions}

Since the discovery of quantum Hall effect \cite{PhysRevLett.45.494,
PhysRevLett.49.405, PhysRevB.23.5632}, physicists have established a new
concept to characterize an exotic set of materials, called topological
insulator, in physics and material science \cite{PhysRevB.74.195312,
Bernevig15122006, RevModPhys.82.3045} . In such insulators, the
electronic bands below the Fermi energy have topological nontrivial
structures which are specified by the numbers called topological
indices \cite{PhysRevB.78.195125}. In experiments, several classes of
materials have been proved to be topological insulators. But the
proposed quantized transmission spectrum in these materials are seldom
reported because the transport though edges is usually rumored by
the finite conductance from the bulk states \cite{Spanton2014,
PhysRevLett.114.096802, Chang12042013}. 

It will be plausible to find similar topological phenomena in systems
``cleaner'' than the electronic ones. There are both theoretical and
experimental works to extend the theory to photonic crystallines
\cite{Fang2012a, Bliokh2015, Wu2015c, Ma2015,
Nalitov2015,PhysRevLett.115.253901,PhysRevB.93.075110}, phonon in solids
\cite{Zhang2010, Wang2009}, exciton systems \cite{Yu2015b,Karzig2015},
electric circuits \cite{Ningyuan2015, Albert2015}, harmonic vibrational
lattices \cite{Wang2015j,1367-2630-17-7-073031,Kariyado2015a,Nash2015,
Roman}, floquet classical system \cite{PhysRevB.93.085105}, and so on.
The last three categories are based on classical mechanics and do not
pose difficulties on the nano-sized material science and may allow
observers to watch the chiral or the helical edge states by eyes
\cite{Nash2015,Roman}.

However the effect of unavoidable dissipations in the classical models
has not been systematically investigated yet. In this paper, we try to
solve this problem based on a 2-dimensional (2D) topological mechanical
lattice.  This investigation extends our understanding on these systems
in two aspects. The first one is that our discussions are not based on a
Hamiltonian representation because in fact there is no Hamiltonian for
such systems. So it is impossible to borrow the discussions on the
language of Hamiltonian from its quantum counterpart directly. The
second one is that the symmetry which protects edge states should be
reexamined in the presence of the dissipation. It is believed that the
dissipation will induce decoherence and dephasing in quantum
systems\cite{PhysRevLett.112.133902,PhysRevLett.102.065703,PhysRevLett.115.040402,
PhysB84,naturep7, njp15, Bardyn2}. For instance, in the quantum Hall
system, it will fuzz up the universal values of the Hall plateaus. On
the contrary, we will show that in the classical systems the dissipation
will not break the gauge invariance. This makes the chiral edge states
robust against the strong dissipation. We will perform both gauge
argument and numerical calculation to confirm this conclusion.

Furthermore, we find that the dissipation will induce a damping
spectrum. Such damping dispersion can generate an effective force for a
free moving wave packet, which seems to break the Newton's first law at
the first glance. We can understand this effect by considering a free
wave packet whose profile in the wave vector space is a Gaussian
function, $G(\vec q) \sim e^{-\alpha (\vec q - \vec q_0)^2}$.  In a
dissipativeless system without external force, this profile does not
alter with time so that the averaged wave vector $\langle \vec q
\rangle$ is a constant, $\frac{d \langle q \rangle}{dt}=0$. But in the
dissipative system, the damping rate is a function of wave vector $\vec
q$, called the dispersion of damping.  As a result, the profile is
varying with time although the wave vector $q$ is still good quantum
number. Effectively, this variance can be expressed as an extra force
acting on the wave packet, $\frac{d \langle q \rangle}{dt}= F(\vec
q_0)$. Such force and the nonzero Berry curvature will bend the trace of
the wave packet, which can be observed in experiments.

The paper is organized as the following: In section 2, we present a
topological mechanical system on a rotating square lattice. All our
arguments will base on this model. In section 3, we introduce how to
calculate the Chern number by the evolution of the centers of the
Wannier functions. The gauge invariance that protecting the chiral edge
states is discussed. We further illustrate that such gauge argument
is still available in a disordered lattice.
In section 4, we derive the equation of motion for a free wave packet.
The dissipation induced force is discussed. Section 5 is for the
conclusions and outlooks.

\section*{The model: a rotating square lattice}

\begin{figure}
  \centering
  \includegraphics[width=87mm]{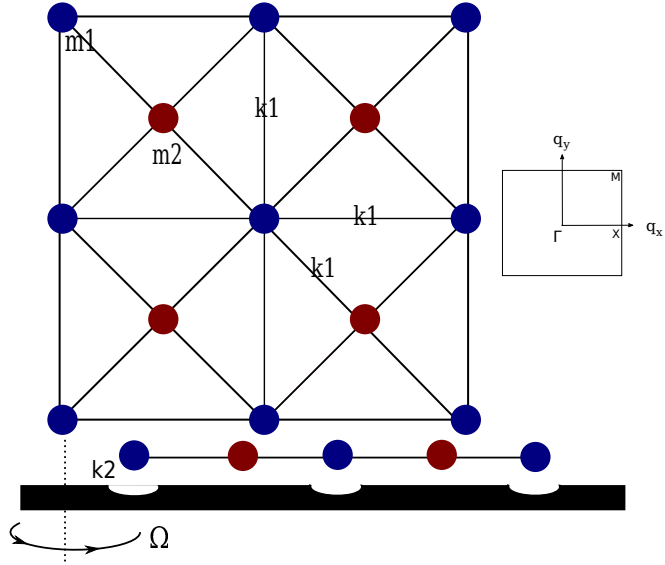}
  \caption{The square phononic lattice composed of MPs
    and springs, with top view (upper panel) and side view (lower panel).
    Each unit cell has two MPs with masses $m_1$ and $m_2$ respectively.
    The dark lines represent the springs with the spring constant $k_1$.
    The lattice is mounted on a
    rotating plane whose angular velocity is $\Omega$. Besides the upward
    supportive forces, the plain also interacts with MPs by restoring
    forces whose effective spring constant is $k_2$ in the horizontal
    directions. We use the pits on the plane to represent such harmonic
    forces. Throughout this paper, we take $m_1=m_2=m$ and $k_2=0.2
    k_1$. The first Brillouin zone is shown on the right. Here
the lattice constant is taken as the length unit.}
  \label{fig1}
\end{figure}

We first consider a square lattice shown in Fig. \ref{fig1}. Each unit
cell has two inequivalent mass points (MPs) and each MP has two degrees
of freedom, $R_x$ and $R_y$ in the plane. After Fourier transformation
to the wave vector space $\vec q$, the vibrational motions are determined
by such Newton's equations:
\begin{eqnarray}
  M \ddot{ \delta\vec{ R}} =K \delta\vec{R} + G \dot{\delta\vec{R}}
  +\Gamma \dot{\delta\vec{R}}.
  \label{eq1}
\end{eqnarray}
Here $\delta \vec{R}= (\delta x^{(1)}, \delta y^{(1)}, \delta x^{(2)},
\delta y^{(2)})^T$ is the vector with the components representing the
displacements of MPs, 1 and 2, away from their rest positions along
$x$ and $y$ directions for the Bloch wave. The three terms on the right
hand side stand for the restoring forces, the Coriolis forces and the
dissipative forces, respectively. 

The matrix $M$ is
\begin{equation}
  M=\begin{pmatrix} m_1 & 0 & 0 & 0 \\ 0 & m_1 & 0 & 0 \\ 0 & 0 & m_2
    & 0 \\ 0 & 0 & 0 & m_2 \end{pmatrix},
\end{equation}
with $m_1$ and $m_2$ are the masses of the two MPs in each unit cell. 
The restoring force matrix $K$ reads
\begin{widetext}
\begin{equation}
  K =2\begin{pmatrix} -k_1 \sin^2(\frac{q_x}{2}) -\frac{k_2}{2} & 0 & 0 & 0 \\
    0 & -k_1 \sin^2(\frac{q_y}{2}) -\frac{k_2}{2}& 0 & 0 \\
    0 & 0 & 0 & 0 \\
    0 & 0 & 0 & 0 \end{pmatrix} + 
  \frac{k_1}{2} \begin{pmatrix} -4 & 0 &
    \gamma_+(q_x,q_y) & \gamma_-(q_x,q_y) \\
    0 & -4 & \gamma_-(q_x,q_y) &
    \gamma_+(q_x,q_y) \\
    \gamma_+^*(q_x,q_y) &
    \gamma_-^*(q_x,q_y) & -4 & 0 \\
    \gamma_-^*(q_x,q_y) &
    \gamma_+^*(q_x,q_y) & 0 & -4 \end{pmatrix},
\end{equation}
\end{widetext}
where $k_1$ and $k_2$  are the spring constants shown in Fig.
\ref{fig1}, $\gamma_+(q_x,q_y)= 1+e^{iq_x}+e^{iq_y}+e^{i(q_x+q_y)}$ and
$\gamma_-(q_x,q_y)=1- e^{iq_x} - e^{iq_y} +e^{i(q_x+q_y)}$, and
$q_x$,$q_y$ are the wave vectors of the phonon modes
along $x$ and $y$ directions, respectively. It is derived from the
harmonic potential: $V= \sum_{ij} \frac{1}{2} k_2 (\delta
{x^{(1)}_{ij}}^2 + \delta {y^{(1)}_{ij}}^2) + \frac{1}{2} k_1 [(\delta
x^{(1)}_{ij} -\delta x^{(1)}_{i+1j})^2 + (\delta y^{(1)}_{ij}- \delta
x^{(1)}_{ij+1})^2] + \frac{1}{4}k_1 \{ [(\delta x^{(2)}_{ij}+ \delta
y^{(2)}_{ij})-(\delta x^{(1)}_{ij}+\delta y^{(1)}_{ij})]^2 + [(\delta
x^{(2)}_{ij}+ \delta y^{(2)}_{ij}) - (\delta x^{(1)}_{i+1j+1}+\delta
y^{(1)}_{i+1j+1})]^2 + [(\delta x^{(2)}_{ij}- \delta y^{(2)}_{ij}) -
(\delta x^{(1)}_{i+1j}- \delta y^{(1)}_{i+1j})]^2 + [(\delta
x^{(2)}_{ij}- \delta y^{(2)}_{ij}) - (\delta x^{(1)}_{ij+1}- \delta
y^{(1)}_{ij+1})]^2 \}$, where the subscripts $i$ and $j$ label the
positions of a unit cell in the $x$ and $y$ directions.   

$G$ is the matrix describing the Coriolis force acting on the lattice,
\begin{equation}
  G=2 \begin{pmatrix} 0 & m_{1} \Omega & 0 & 0 \\ -m_1 \Omega & 0 & 0 &
    0 \\ 0 & 0 & 0 & m_2 \Omega  \\ 0 & 0 & -m_2 \Omega & 0 \end{pmatrix},
\end{equation}
where $\Omega$ is the angular velocity of the plane with the positive
direction defined in Fig. \ref{fig1}. The stationary centrifugal forces do not
appear in this equation because we have taken the variations $\delta
\vec R$ with respect to their new stationary positions where the centrifugal
forces are compensated by the restoring forces of the springs. One may
worry about the transverse motion because the spring is stretched in
this case. But as the whole lattice is placed abaxial and the lattice
constant is small, each mass
point will fell similar centrifugal force. In this case, some of strings
are stretched and some are compressed. So totally, they produce zero net
force in the transverse direction. The
centrifugal forces will also induce forces in the radial direction when the MPs
are away from their new stationary positions. We also ignore them in the
case of small $\Omega$ because
their magnitudes are proportional to $\Omega^2$.
For the dissipation term, we take isotropic dissipative forces with 
$\Gamma=\gamma I$, where $I$ is a $4\times 4$ unit matrix.

Such second order differential equations can be solved as an eigenvalue
problem by introducing new variables, $\vec p = M
\dot{\delta\vec{R}} - G \delta\vec{R} -\Gamma \delta\vec{R}$. The Eq.
\ref{eq1} is rewritten  as
\begin{equation}
  \begin{pmatrix} \dot{\delta\vec{R}} \\ \dot{\vec{p}} \end{pmatrix} =
  \begin{pmatrix} M^{-1}(G+\Gamma) & M^{-1} \\  K & 0 \end{pmatrix} \begin{pmatrix}
    \delta\vec{R} \\ \vec{p} \end{pmatrix} = A \begin{pmatrix}
        \delta\vec{R} \\ \vec{p} \end{pmatrix}.
  \label{eq2}
\end{equation}
As $\delta\vec{R}$, as well as $\vec p$, is varying with time as $\sim
e^{i\omega t}$ for an eigenstate, the phonon dispersion will be obtained by solving  the
eigenvalue problem: $det(i\omega I -A)=0$, where
$A$ is the matrix on the right hand side of Eq. \ref{eq2}.

\begin{figure}[ht]
  \centering
  \includegraphics[width=87mm]{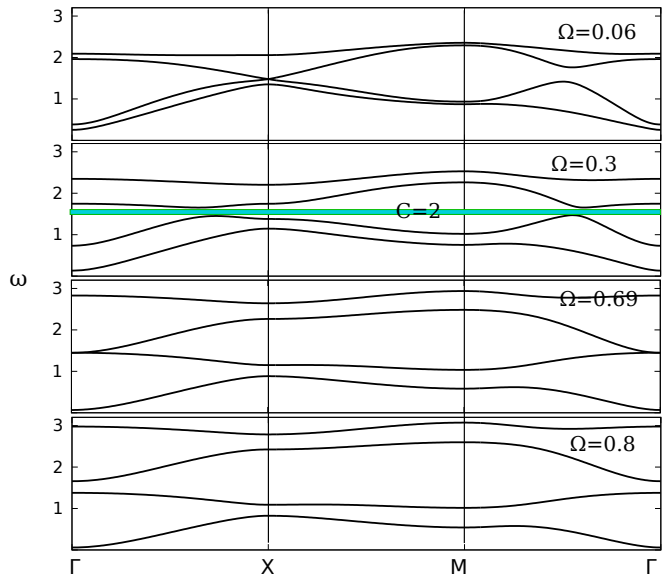}
  \caption{The phonon dispersion for the dissipativeless model. The
  angular velocity is taken at $\Omega=0.06$, $0.3$, $0.69$ and $0.8$
in the panels from up to down. The model is in a topological phase
when $0.06 < \Omega < 0.69$. We illustrate the topologically nontrivial
gap with a green box and indicate its Chern number $C=2$ there
explicitly. All other gaps are topological trivial.}
  \label{fig2}
\end{figure}

In the absence of dissipation, the lattice has a mechanical topological
phase. In Fig. \ref{fig2}, we show the phonon dispersion at several
$\Omega$. The parameters are taken as $m_1=m_2=m$, $k_2=0.2 k_1$ and
$\gamma=0$. Throughout this paper, all physical quantities are taken as
dimensionless. It is easy to recast the unit at the end of
the calculations (for instance the eigen-frequencies $\omega$ and the
angular velocity $\Omega$ are both taken in the units of $\sqrt{k_1/m}$). 
This model has a topologically nontrivial phase when $0.06 <
\Omega < 0.69$. We explicitly denote the topologically nontrivial gap
corresponding to a Chern number, $C=2$, by the green box in the figure.
All other gaps are topologically trivial. 

\section*{The Chern number in the present of dissipation}

Now we discuss how to calculate the Chern number in the presence of
dissipation.  At the very beginning, a definition of inner product for
the eigenstates must be discussed. It has been noticed that although the
topological mechanism is considered as a classical mimic of the Chern
insulator, there is still a technical difference between them. In the
last section, we have shown that the motion of the classical lattice follows
the Newton's law and the vibrational eigenstates are obtained from a
non-hermitian matrix $A$.  So the eigenstates with different eigenvalues
are not orthogonal to each other as usual. Such technical problem has
been solved by a redefinition of inner product for the eigenstates
\cite{Wang2009,Wang2015j}. But the problem becomes worse as the
dissipation is included in the system. The inner product in Ref.
\onlinecite{Wang2009}, $\epsilon^\dagger \epsilon + \frac{i}{\omega}
\epsilon^\dagger A \epsilon$, fails because there is no Hamiltonian for
a dissipative system. While the definition of inner product in Ref.
\onlinecite{Wang2015j},$\epsilon \tilde{M} \epsilon$, becomes
non-orthogonal again. Here $\epsilon$ is the polarization vector, $A$ is
the effective vector potential in the Hamiltonian and $\tilde{M}$
is the effective mass matrix. In this paper, the inner product of the
eigenstates is redefined by the left and right eigenstates directly. We
denote the state in the bra-ket notation with $\langle u_a|$ and $|u_a
\rangle$, which are the left and the right $a$th eigenstates of the
matrix $A$ respectively. For the case $a\ne b$, it is easy to find that
$\langle u_a|u_b \rangle=0$. One should note that, in the new
definition, the states include both the components of the position
variables $\delta \vec R$ and the auxiliary variables $\vec p$. 

Now we describe how to calculate the Chern number by studying the
evolution of the centers of Wannier functions. It is extended
from the numerical method developed for electronic topological
insulators. Here we first briefly review this method in the electronic system. 
The Chern number is expressed as \cite{PhysRevB.84.075119, PhysRevB.83.235401},
\begin{eqnarray}
C = &\sum_a \frac{1}{2\pi} \oint \langle
u_{ak} | i\frac{\partial}{\partial k} | u_{ak} \rangle \cdot dk
\nonumber \\
 = & \sum_a
\frac{1}{2\pi} \int dk_y \frac{d \int d k_x \langle
  u_{ak} | i\frac{\partial}{\partial k_x} | u_{ak} \rangle}{d k_y},
\end{eqnarray}  
where the summation is over all the occupied bands and the closed path
integration is along the boundaries of the first Brillouin zone (BZ). As $i
\frac{\partial}{\partial k_x}$ is the position operator in the $x$
direction, the Chern number can be written down as $C=\sum_a \int d k_y
\frac{dX_a}{d k_y}$, where $X_a = \frac{1}{2\pi} \int d k_x \langle
u_{ak} | i\frac{\partial}{\partial k_x} | u_{ak} \rangle$ is the
position ( in the $x$ direction) of the Wannier functions for the $a$th
band.  Yu et al. have developed a numerical method to calculate  $e^{i2
\pi X_a}$ instead of $X_a$\cite{PhysRevB.84.075119}. Its physical
meaning is to calculate the accumulated phase for the $a$th band when
changing $k_x$ by one reciprocal vector in the BZ. One can divide such
variation into $M$ pieces, and each piece is a nonzero matrix only for
$\langle k_x+\frac{2\pi}{M}| e^{i \frac{2\pi}{M} \hat X} |k_x \rangle$,
in the momentum space. $e^{i2 \pi X_a}$ will be obtained from the
eigenvalues of a product matrix by multiplying all these nonzero
matrices. This numerical method has been applied to investigate the
topologically nontrivial electronic bands \cite{ Taherinejad2014,
ye20151, Ezaki2015}.

In our classical system, the Chern number can still be considered as the
total displacement of the Wannier functions (in the $x$ direction) for
the studied phonon bands as changing $q_y$ by one reciprocal vector. But
the inner product for $|u_{aq}\rangle$ and $\langle u_{aq}|$ must take
the form of the new definition.  As a result, the project operator
inserted into the pieces $e^{i \frac{2\pi}{M} \hat x}$ should be changed
to $\sum_a \frac{| u_{aq} \rangle \langle u_{aq} |}{\langle u_{aq} |
u_{aq} \rangle} $. So the total phase $\Phi(q_y)$ for the interesting
$n$ phonon bands is $\Phi=\sum_{a=1}^n arg[\beta_a]$, where $\beta_a$ is
the eigenvalues of the $n\times n$ matrix $B=\prod_{i=1}^M B^i$ with 
\begin{equation}
  B^i_{ab} = 
  \begin{cases}
    \langle u_{a,q_{x}=2\pi; q_y}| u_{b,q_x=2\pi\frac{M-1}{M};q_y} \rangle
    \quad
    i=M , \\
    \frac{\langle u_{a,q_x=\frac{2\pi i }{M};q_y}  | u_{b,
      q_x=\frac{2\pi(i-1)}{M};q_y} \rangle}{\langle u_{a,q_x=\frac{2\pi i}{M};
    q_y} | u_{a,q_x=\frac{2\pi i}{M};q_y} \rangle} \quad  i=1,2,\dots,M-1.
    \end{cases}
\end{equation}
Here $|u_{a,q_x;q_y} \rangle$ and $\langle u_{a,q_{x};q_y}|$ are the right and left
$a$th eigenvectors of the matrix $A$ in Eq. \ref{eq2} and $n$ is the
number of bands below the interesting gap. 

In this model, because the topologically nontrivial gap is in between
the second and the third phonon bands, the matrix $B^i$ in the above
expression is a $2\times 2$ matrix with $a$ and $b$ run over the indeces
of phonon bands, 1 and 2. 

\begin{figure}[ht]
  \centering
  \includegraphics[width=87mm]{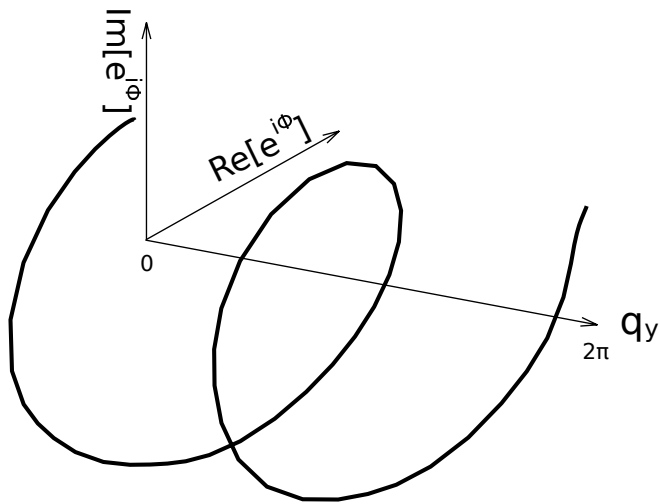}
  \caption{The total phase $\Phi$ in  $e^{i\Phi}= e^{i 2\pi \sum_a X_a}$ as
$q_y$ is varying $2\pi$. The angular velocity is $\Omega=0.3$ and the
strength of dissipation is $\gamma =0.6$. $e^{i 2\pi \hat X}$ is divided into $M=10$ pieces in the
calculation.}
  \label{fig3}
\end{figure}

In Fig. \ref{fig3}, we show how the total phase in $e^{i 2\pi \sum_a
X_a}$ changes as $q_y$ is varying $2\pi$. Here the angular speed is
$\Omega=0.3$ and the dissipative strength is $\gamma=0.6$. We see that
the Wannier function centers move totally $2$ unit cells as $q_y$ is
varying one reciprocal vector. So we can conclude that the Chern number $C$
is $2$ for the gap between the second and the third bands.

As we expected, dissipation will not alter the Chern number as long as
the band gap is not closed. Now we employ a $N_x\times N_y$ lattice in
the cylinder geometry and discuss the inevitable chiral edge states just
like Laughlin did in quantum Hall system \cite{PhysRevB.23.5632}.  In
such a classical system, there is neither Fermi energy nor Fermi-Dirac
distribution for the eigenstates. But as the eigenvalue spectrum is an
intrinsic property and is independent of the number of the excitations
in the system, we can assume a special situation in which all the
eigenstates below the gap are excited and the states above that are
empty. Now supposing there is a gauge transformation that continously
changing $q_y \to q_y + \frac{2\pi}{N_y}$, for the wave vector in the
azimuthal direction.  As the states in an individual band are all
excited or all empty for $q_y$s, such gauge transformation is equivalent
to a change of one $q_y$ by $2\pi$.  From the evolution of the Wannier
centers in the cylindrical direction as varying $q_y$, two excited
Wannier states have been pumped from the left edge to the right edge.
This gauge transformation is imitating the magnetic flux quanta
penetrating the center of a cylinder in electronic quantum Hall system.
As the variance of $q_y$ by one reciprocal vector changes nothing, the
gauge transformation is invariant and the system must come back to its
initial state after the transformation. To compensate the pumped bulk
states, there must have edge states at the boundaries of the cylinder.
Such edge states will connect the low excited bands to the high empty
bands twice during the pump. According to this argument, the existence
of the chiral gapless edge states in the spectrum is concluded and such
edge states must be robust against strong dissipation.

\begin{figure}[ht]
  \centering
  \includegraphics[width=148mm]{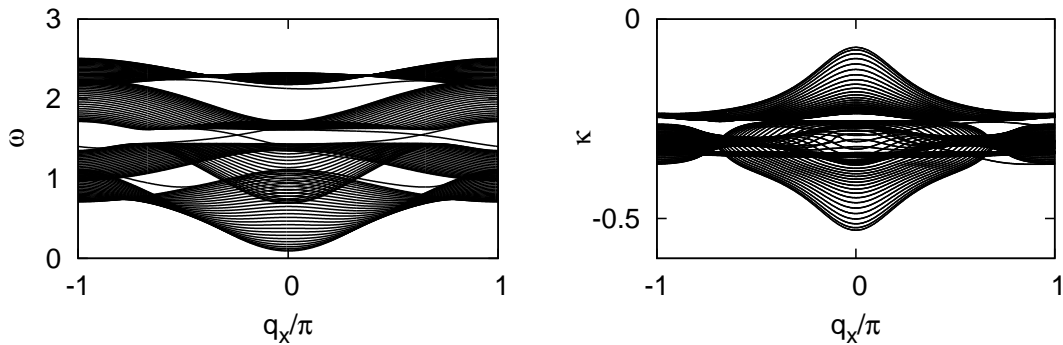}
  \caption{The real parts (left) and the imaginary parts (right) of the eigen-frequencies 
    for a ribbon with width $W=30$. 
  The two chiral edge modes with finite life time are confirmed.} 
  \label{fig4}
\end{figure}

We confirm the above conclusion by a calculation of spectrum of a ribbon
with two geometric boundaries.  In the left panel of Fig. \ref{fig4}, we
show the real parts of the eigenvalues of the ribbon with width $W=30$.
All parameters are the same as those in Fig. \ref{fig3}. An open boundary
condition is taken in the transverse direction and the wave vector $q_x$
along the longitudinal direction is a good quantum number.  We can see
two chiral edge modes whose eigen-frequencies are within the gap between
the second and the third phonon bands. Interestingly, this gap is more pronounced in the
dissipative case than that in the dissipativeless case. The dissipation
induced damping, which corresponds to the imaginary parts of the
eigenvalues, are shown in the right panel of Fig. \ref{fig4}.  It can be
seen that all eigenstates are damped, but the damping rate is not
uniform for different eigenstates at different $q_x$. This is the origin
of the dissipation induced force that is discussed in the next section.

\begin{figure}[ht]
  \centering
  \includegraphics[width=87mm]{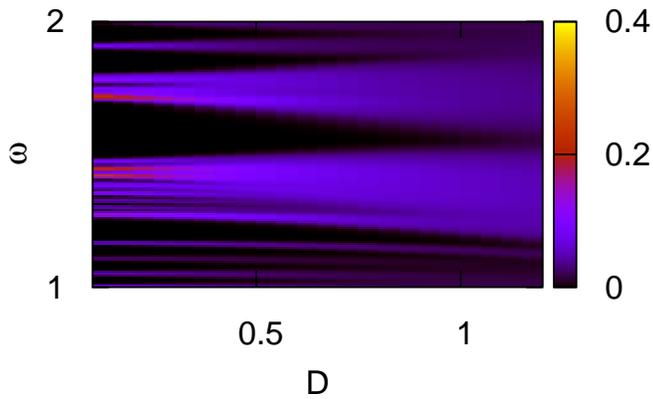}
  \caption{The averaged density of states (in arbitrary unit) as a function of the strength of disorder
  $D$ and the frequency $\omega$. The gap (at $\omega \sim 1.5$) 
  between the bands 2 and 3 is closed around $D=0.9$.}
  \label{fig5}
\end{figure}

At the end of this section, we show that the above gauge argument can be
extended to the disordered case. In Fig.  \ref{fig5}, we show the
averaged density of states for the disordered $N\times N$ phononic
lattice. Here $N=10$ and the average is over $2000$ samples. The spring
constant $k_1$ is randomly chosen in the range
$[(1-D/2)k_1,(1+D/2)k_1]$. Other parameters are fixed and are the same
as those in Fig. \ref{fig3}. It is found that the gap between the second
band and the third band is not closed until the strength of disorder
reaches $D=0.9$. When the gap is not closed, the gauge argument is
performed on an ensemble of cylinders with different disordered
configurations.  The number of pumped bulk states can still be
investigated by the evolution of the centers of the Wannier functions.
Similar to that in the electronic system \cite{Niu1985}, this evolution
can be calculated with the present method on the ensemble of
super-lattices in which $N\times N$ disordered samples are taken as the
unit cell. In this case, the first and the second bands split into
totally $2N^2$ bands and the total phase $\Phi$ is calculated for these
bands. We find that the Chern number $C$, which characterizes the
evolution of the Wannier functions, is fixed at $2$ in the ensemble. It
will fluctuate from sample to sample only when the gap is closed by the
disorder for $D>0.9$. So from this gauge investigation, it is concluded
that the chiral edge states in the topological mechanical system are
robust against disorder and dissipation. This provides further support
for the conclusions of Fig. 3 in Ref. \onlinecite{Wang2015j}, where the
topological propagation immune against disorder is shown. We have also
studied the randomness of the mass instead of the strength of springs
and found similar results.

\section*{Dissipation induced force and its effect on the motion of wave
packet}

Supposed that there is a free wave packet composed by the states in the
$a$th phonon band. Its profile at the initial time $t=0$ is $e^{-\alpha
|\vec q-\vec q_0|^2}$ in the wave vector space. This initial state is similar to
the states in the discussion of semi-classical equation of motion for
quasiparticles in solid.   

As there is no external force, $\vec q$ is a good quantum number. We can
write down the wave packet in the real space at any time $t$ easily,
\begin{equation}
  | \psi_{a,k_0} (\vec x,t) \rangle \propto \int dq e^{-\alpha |\vec q - \vec
  q_0|^2} e^{i (\vec q \cdot \vec x- \omega^a_q t)} e^{\kappa^a_q t} |
  u_{a,q} \rangle,
  \label{eqwp}
\end{equation}
where $\omega^a_q$ and $\kappa^a_q$ are the real and imaginary parts of
the eigen-frequency for the $a$th eigenstate $|u_{a,q} \rangle$ at the
wave vector $\vec q$. After expanding $\omega^a_q$, $\kappa^a_q$ and $|u_{a,q}
\rangle$ to the first order of the derivation of $\vec q$ at $\vec
q_0$, we rewrite the above equation as
\begin{equation}
  | \psi_{a,k_0} (\vec x,t) \rangle \propto e^{i(\vec q_0 \cdot \vec x -
    \omega_{q_0} t)+\kappa^a_{q_0}t} \int dq e^{-\alpha |\vec q - \vec q_0|^2} e^{ \vec
  \nabla \kappa^a_q \cdot (\vec q -\vec q_0)t}  e^{-i \vec 
  \nabla \omega^a_q \cdot (\vec q -\vec q_0) t} [ | u_{a,q_0} \rangle +
  \vec \nabla | u_{a,q} \rangle \cdot (\vec q -\vec q_0)] e^{i (\vec q -
  \vec q_0) \cdot \vec x}.
  \label{eqwp2}
\end{equation}
One can simplify $[ | u_{a,q_0} \rangle + \vec \nabla | u_{a,q}
\rangle ]$ with the berry phase $\vec \beta_a=\frac{\langle u_{a,q}| \vec
\nabla | u_{a,q} \rangle|_{q=q_0}}{i\langle u_{a,q_0}|
u_{a,q_0}\rangle}$, to $e^{i \vec \beta_a \cdot (\vec q -\vec q_0)}
|u_{a,q_0} \rangle $. After substituting $\vec q_0' = \vec q_0 + \vec \nabla
\kappa^a_q|_{q=q_0}\frac{t}{2\alpha}$ and $\vec x' = \vec x - \vec \nabla \omega^a_q|_{q_0'} t
+\vec \beta_{q_0'}- \vec \nabla \times \vec \beta|_{q_0'} \times \vec \nabla
\kappa^a_q|_{q=q_0'}\frac{t}{2\alpha}$ into the integration and
assuming that $t$ is within a small time interval, the wave packet becomes
\begin{equation}
   | \psi_{a,k_0} (\vec x,t) \rangle \propto \int dq e^{-\alpha |\vec q
   -\vec q_0'|^2} e^{i (\vec q -\vec q_0') \cdot \vec x'}.
  \label{}
\end{equation}
So the motion of wave packet satisfies the following equations of motion
for its centers $\vec Q$ and $\vec X$ in the wave vector and real
spaces, respectively:
\begin{eqnarray}
  \frac{d \vec X^a}{dt}= \vec \nabla \omega^a +
  \vec \Theta^a \times \frac{d\vec Q^a}{dt} ,\\
  \frac{d \vec Q^a}{dt}= \vec \nabla \kappa^a \frac{1}{2\alpha},
  \label{}
\end{eqnarray}
where $\vec \Theta^a$ is the berry curvature. Compared with those of
the dissipationless case, we find that the dissipation will induce an effective
force whose amplitude is proportional to the gradient of the damping
rate and inversely proportional to the size of the wave packet in the
real space.

\begin{figure}[ht]
  \centering
  \includegraphics[width=87mm]{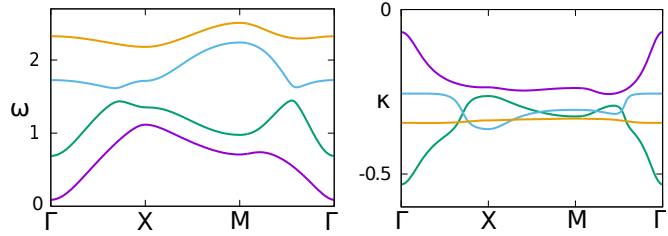}
  \caption{The dispersion of phonon eigen-frequency $\omega$ (left) and damping
  rate $\kappa$ (right) along the high symmetric directions in BZ.}
  \label{fig6}
\end{figure}

In Fig. \ref{fig6}, we show the dispersions of the real and imaginary
parts, $\omega$ and $\kappa$, of the eigen-frequencies for the
dissipative model. It seems that the damping rate is strongly dispersed
in this topological system. This dispersion comes partially from the
finite value of $\Omega$. It is also caused by the fact that the damping
of a vibrational mode is proportional to its velocity.  As the time
derivation of the displacement of an eigen-mode will generate a
multiplier, the eigen-frequency $\omega$, to the velocity, the
eigen-modes with different eigen-frequencies will suffer different
damping rates.

\begin{figure}[ht]
  \centering
  \includegraphics[width=87mm]{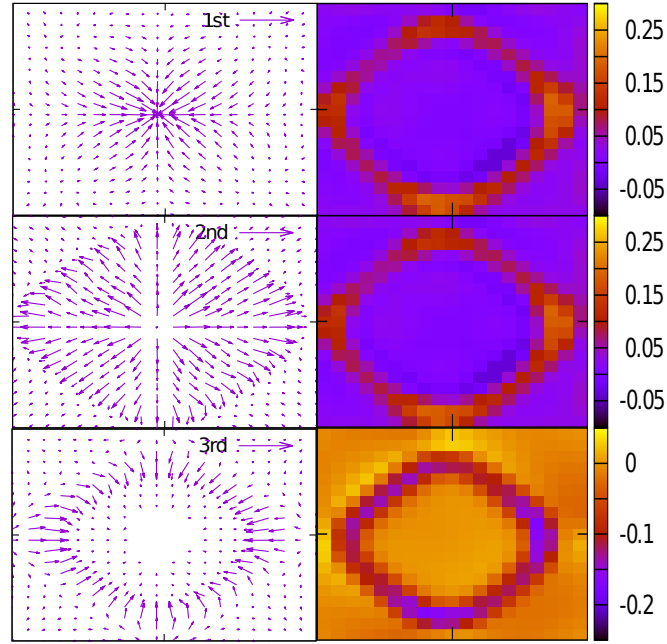}
  \caption{$\vec \nabla \kappa^a$ (left) and the strength of berry curvature $\vec
    \Theta^a$ perpendicular to the BZ plane (right) are shown for the
    first three bands, $a=1,2,3$ from up to down.}
  \label{fig7}
\end{figure}

In Fig. \ref{fig7}, we plot the gradient of $\kappa$, $\vec \nabla
\kappa^a$ in the BZ with small arrows in the left panels and the
strength of Berry curvature $\vec \Theta^a$ with the contour in the
right panels.  The Berry curvature $\vec \Theta^a$ is calculated by
discretizing the BZ into small grids and counting the accumulated
phase by the eigenstates along the edges of each grid \cite{Wang2015j}.
Of course, the new definition of inner product must be taken in the
calculation.  Here only those for the first three phonon bands are
plotted and the forth band is ignored because, as shown in Fig.
\ref{fig6}, $\kappa$ is not varying rapidly for this band. It is shown
that for the 2nd and the 3rd bands, there is a cirque in the BZ, in
which both $\vec \nabla \kappa^a$ and $\vec \Theta^a$ are relatively
large. According to the equations of motion, $ \frac{d \vec X^a}{dt}=
\vec \nabla \omega^a +\frac{1}{2\alpha} \vec \Theta^a \times \vec \nabla
\kappa^a$, the free wave packet composed by the eigenstates of the
these bands suffers a tendency towards the transverse direction caused
by the combination of Berry curvature and the extra force in the last
term. As this term is also varying with $\vec q$, as well as with the
time, the trace of the wave packet is a curved line, similar to the
trace of electron in 2D hall bar with longitudinal electric field. We
suggest that such effect may be observed by preparing the proper wave
packet composited by the states where both $\vec \nabla \kappa^a$ and
$\vec \Theta^a$ are large in the 2nd band or in the 3rd bands.

\section*{Conclusions and outlooks}

The chiral edge states in the topological mechanical system are robust
against dissipation and disorder.  We also find that the dissipation can
induce a dispersion of damping for the eigenstates. As a result, an
extra force appears in describing the motion of free wave packet. After
taking into account the non-zero Berry curvature in such system, we find
that the free wave packet will swerve even through there is no external
force acting on the packet. One should be aware that the Coriolis forces
and dissipative forces have been taken into account during the
calculation of bands so that they should not be considered as external
forces again during the calculation for the wave packet.  We also
suggest that such effect should also exist in dissipative photonic
system.  Besides this, a general method to calculate the Chern numbers
in such systems is developed and we believe that it is also applicable
in non-hermitian quantum systems.  

{\it
Acknowledgments.---} The work was supported by 
National Foundation of Natural Science in China Grant Nos. 10704040,
11175087.

{\it Author contributions statements:}
Ye Xiong wrote the paper and Tianxiang Wang prepared figure 1. All
authors reviewed the manuscript.

{\it Additional information}

Competing financial interests:
The authors, Ye Xiong, Tianxiang Wang and Peiqing Tong, declare no competing financial interests.


\end{document}